\documentclass[runningheads]{svmult}

\usepackage{makeidx}   
\usepackage{graphicx}  
\usepackage{subeqnar}  
\usepackage{multicol}  
\usepackage{physprbb}  
\makeindex             

\newcommand{\expect}[1]{\langle #1 \rangle}
\newcommand{\abs}[1]{\left| #1 \right|}

\newcommand{\be}{\begin{equation}}
\newcommand{\ee}{\end{equation}}
\newcommand{\kvec}{{\bf k}}
\newcommand{\rvec}{{\bf r}}

\newcommand{\xvec}{{\bf x}}
\newcommand{\bea}{\begin{eqnarray}}
\newcommand{\eea}{\end{eqnarray}}
\begin{document}
\title*{The Coupled Electronic-Ionic Monte Carlo Simulation Method}
\toctitle{The Coupled Electronic-Ionic \protect\newline Monte
Carlo Simulation Method }
%
%
\titlerunning{Coupled Electronic-Ionic Monte Carlo}
%
\author{David Ceperley\inst{1,2}
\and Mark Dewing\inst{3} \and Carlo Pierleoni\inst{1,4}}
\authorrunning{Ceperley, Dewing, Pierleoni}
%
%
\institute{CECAM, c/o ENS Lyon, 46 All\'ee d'Italie, 69364 Lyon
(France) \and Department of Physics, University of Illinois at
Urbana-Champaign, 1110 West Green Street, Urbana, Illinois 61801
(USA) \and KSL, Intel Corp., 1906 Fox Dr., Champaign, IL, 61820
USA \and INFM and Department of Physics, University of L'Aquila,
Via Vetoio, L'Aquila (Italy)}

\maketitle              

\begin{abstract}
\index{abstract} Quantum Monte Carlo (QMC) methods such as
Variational Monte Carlo, Diffusion Monte Carlo or Path Integral
Monte Carlo are the most accurate and general methods for
computing total electronic energies. We will review methods we
have developed to perform  QMC for the electrons coupled to a
classical Monte Carlo simulation of the ions. In this method, one
estimates the Born-Oppenheimer energy $E(Z)$ where $Z$ represents
the ionic degrees of freedom. That estimate of the energy is used
in a Metropolis simulation of the ionic degrees of freedom.
Important aspects of this method are how to deal with the noise,
which QMC method and which trial function to use, how to deal with
generalized boundary conditions on the wave function so as to
reduce the finite size effects. We discuss some advantages of the
CEIMC method concerning how the quantum effects of the ionic
degrees of freedom can be included and how the boundary conditions
can be integrated over. Using these methods, we have performed
simulations of liquid H$_2$ and metallic H on a parallel computer.
\end{abstract}

\section{Introduction}
The first computer simulations of a condensed matter system used
the simplest inter-atomic potential, the hard sphere interaction\cite{metropolis53}.
 As computers and simulation methods progressed, more
sophisticated and realistic potentials came into use, for example
the Lennard--Jones potential to describe rare gas systems, the
potential functions being parameterized and then fit to reproduce
experimental quantities. Both Molecular Dynamics (MD) and Monte
Carlo (MC) methods can be used to generate ensemble averages of
many-particle systems, MC being simpler and only useful for
equilibrium properties.

Inter--atomic potentials originate from the microscopic structure
of matter, described in terms of electrons, nuclei, and the
Schr\"odinger equation. But the many-body Schr\"odinger equation
is too difficult to solve directly, so approximations are needed.
In practice, one usually makes the one electron approximation,
where a single electron interacts with the potential due to the
nuclear charge and with the mean electric field generated by all
the other electrons. This is done by Hartree--Fock (HF) or with
Density Functional Theory (DFT)\cite{parr89}. DFT is, in
principle, exact, but contains an unknown exchange and correlation
functional that must be approximated,  the most simplest being the
Local Density Approximation (LDA) but various improvements are
also used.

In 1985, Car and Parrinello introduced their method, which
replaced an assumed functional form for the potential with a
LDA-DFT calculation done ``on the fly''\cite{car85}. They did a
molecular dynamics simulation of the nuclei of liquid silicon by
computing the density functional forces of the electronic degrees
of freedom at every MD step. It has been a very successful method,
with the original paper being cited thousands of times since its
publication. There are many applications and extensions of the
Car--Parrinello method\cite{payne92,marx96,sprik00,tuckerman00}.
The review of applications to liquid state problems by
Sprik\cite{sprik00} notes that the LDA approximation is not
sufficient for an accurate simulation of water although there are
improved functionals that are much more accurate.

Quantum Monte Carlo (QMC) methods have developed as another means
for accurately solving the many body Schr\"odinger
equation\cite{foulkes,hammond94,anderson95,ceperley96}. The
success of QMC is due largely to the explicit representation of
electrons as particles, so that the electronic exchange and
correlation effects can be directly treated. Particularly within
the LDA, DFT has known difficulties in handling electron
correlation\cite{grossman95}.

In the spirit of the Car-Parrinello method, in this paper we
describe initial attempts to combine a Classical Monte Carlo
simulation of the nuclei with a QMC simulation for the electrons.
This we call Coupled Electronic-Ionic Monte Carlo
(CEIMC)\cite{dewing00b}. As an example of this new method we apply
it to warm dense many-body hydrogen.  Hydrogen is the most
abundant element in the universe, making an understanding of its
properties important, particularly for astrophysical applications.
Models of the interiors of the giant planets depends on a
knowledge of the equation of state of
hydrogen\cite{hubbard84,stevenson88}. Hydrogen is also the
simplest element, but it still displays remarkable variety in its
properties and phase diagram. It has several solid phases at low
temperature, and the crystal structure of one of them (phase III)
is not fully known yet. At high temperature and pressure the fluid
becomes metallic, but the exact nature of the transition is not
known, nor is the melting transition from liquid to solid for
pressures above 1\,MBar. The present knowledge of the phase
diagram of hydrogen is summarized in Figure 1.

\begin{figure}[tb]
\begin{center}
\includegraphics[width=.6\textwidth]{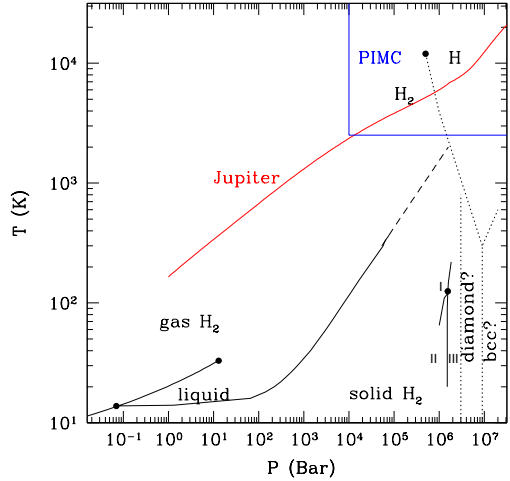}
\end{center}
\caption[]{ Phase Diagram of hydrogen. Solid lines are
experimental determination, dashed line are theoretical estimates.
Red curves show estimates of slices through the giant planets. The
box shows roughly the domain of applicability of PIMC. }
\label{fig:phasediagram}
\end{figure}

Some of the previous QMC calculations have been at high
temperature using the restricted Path Integral MC method. This
method becomes computationally inefficient at temperatures a
factor of ten lower than the Fermi temperature\cite{militzer00a}.
At the present time it is not known how to make the PIMC method
efficient at the low temperatures needed to calculate interesting
portions of the phase diagram. Zero temperature QMC methods have
been used for calculations in the ground
state\cite{ca87,natoli93,natoli95} with full quantum effects used
for both the electronic and protonic degrees of freedom. In such
cases it is hard to ensure that the protonic degrees of freedom
are fully converged because of the problem that the electron and
protons require two different time scales which differ by three
orders of magnitude. In addition, finite temperature effects of
the protons are beyond the reach of the method. CEIMC provides a
middle way\,: the electrons are at zero temperature where accurate
trial functions are known and the zero variance principle applies,
while the protons (either classical or quantum) are at finite
temperature and not subjected to the limitations imposed by the
electronic time scale.

The electrons are assumed to be in their ground state, both in the
Car--Parrinello method and in CEIMC.  There are two internal
effects that could excite the electrons, namely coupling to nuclear
motion and thermal excitations. In the first case, we make the
Born--Oppenheimer approximation, where the nuclei are so much more
massive than the electrons that the electrons are assumed to
respond to nuclear motion instantaneously, and so stay in their
ground state. We neglect any occupation of excited states of the
electrons due to coupling to nuclear motion.  To estimate the
effect of thermal excitation in metallic hydrogen, consider a gas
of degenerate electrons at a density of $n=0.0298$ electrons per
cubic Bohr (i.e. $r_s = \left(4 \pi n/3\right)^{-1/3} = 2.0$).
This has a Fermi temperature of about 140,000\,K. In the molecular
hydrogen phase, the gap between the ground state and the first
excited state of a hydrogen molecule at the equilibrium bond
distance is about 124,000\,K. Since our temperatures are well below
this, and we are not at too high pressures (since the pressure
decreases the gap), the thermal occupation of excited states can
be neglected. At higher pressure however, when the electrons becomes
delocalized and the system becomes metallic thermal effects can be relevant.

This report brings up-to-date previous work on CEIMC described in
ref. \cite{dewing}. The rest of this paper is as follows. First,
we will describe the penalty method to rigorously deal with the
noisy QMC estimates of energy differences.  Then we will briefly
discuss method for computing energy differences. Next, the choice
of trial wave function will be discussed. Finally, we put all the
pieces of a CEIMC simulation together and discuss preliminary
results appropriate to many-body hydrogen.

\section{The Coupled Electronic-Ionic Monte Carlo Method}

First let us recall the basic ideas of Variational Monte Carlo
(VMC) and Diffusion Monte Carlo. VMC uses the Metropolis method to
sample the ratio of integrals and gives an upper bound to the
exact ground state energy.
\be
\label{variationalE}
E=\frac{\int dR \abs{\psi_T(R)}^2 E_L(R)}{\int dR \abs{\psi_T(R)}^2}
\ee
where $E_L = (H \psi_T)/\psi_T$ is the local energy.
Important features of VMC are that any computable trial function
can be used for $\psi_T$ and that the statistical uncertainty
vanishes as $\psi_T$ approaches an exact eigenstate.

The second QMC method we apply is diffusion Monte Carlo (DMC) in
which the Hamiltonian is applied to the VMC distribution to
project out the ground state: \be \phi(t) = \psi_T  \E^{-(H-E_T)t}
\phi(0)/\psi_T. \ee The VMC method, though it can directly include
correlation effects, is not sufficiently accurate, at relevant
temperatures, as we discuss below. The projection is implemented
by a branching, drifting random walk\cite{reynolds82} though there
are some advantages to working in a time independent framework of
ground state path integrals.  To maintain a positive function,
needed for efficient sampling, the fixed-node approximation is
used. Though an uncontrolled approximation, estimates of the
resulting error lead to the conclusion\cite{foulkes} that the
systematic error of this approximation are small, especially when
accurate nodal surfaces are used.

In the CEIMC method we move the protons with a ``classical'' Monte
Carlo  and accept or reject to satisfy detailed balance. The
Metropolis acceptance formula is \be A
=\min\left[1,\exp(-\Delta)\right] \ee where $\Delta = \beta[V(s')
- V(s)]$ and $V(s)$ is the BO electronic energy, computed with one
of the QMC methods. The QMC simulation will yield a noisy estimate
for $\Delta$, which we denote as $\delta$. The exponential in the
acceptance ratio is nonlinear, so that $\expect{\exp(-\delta)}
\neq \exp(\expect{-\delta})$. The noise will introduce a bias into
our acceptance ratio formula. Such bias is unacceptable since the
main motivation for the CEIMC method is to improve the accuracy
beyond what can be achieved with alternative approaches. To avoid
this bias in our simulations, we can either run until the noise is
negligible, but that is {\it very} time-consuming, or we can use
the penalty method\cite{ceperley99} which tolerates noise. We
describe this method next.

\section{The Penalty Method}

The basis of the penalty method is to satisfy detailed balance on
average by using information about the energy differences. We
introduce the ``instantaneous'' acceptance probability,
$a(\delta)$, which is a function of the estimated energy
difference.  The average acceptance probability is the acceptance
probability averaged over the noise,
\be A(s\rightarrow s') =
\int_{-\infty}^{\infty} d\delta P(\delta; s \rightarrow s')
a(\delta).
\ee
We need to satisfy detailed balance on average,
\be
A(s \rightarrow s') = A(s' \rightarrow s)
\exp\left[-\Delta\right]
\ee
If the noise is normally distributed
with variance, $\sigma$, it has the distribution
\be P(\delta)  = (2 \sigma^2 \pi)^{-1/2}
      \exp\left[-\frac{(\delta - \Delta)^2}{2\sigma^2} \right].
\ee
Then  a simple solution that satisfies average detailed
balance is
\be a(\delta) =
\min\left[1,\exp(-\delta-\frac{\sigma^2}{2})\right]
\ee
The extra
$-\sigma^2/2$ term causes additional rejections of trial moves due
to noise. For this reason, it is called the penalty method.

An important issue is to verify that the energy differences are
normally distributed. Recall that if moments of the energy are
bounded, the central limit theorem implies that given enough
samples, the distribution of the mean value will be Gaussian.
Careful attention to the trial function to ensure that the local
energies are well behaved may be needed.

In practice, the variance is also estimated from the data, and a
similar process leads to additional penalty terms. Let $\chi$ be
the estimate for $\sigma$ using $n$ samples. Then the
instantaneous acceptance probability is \be a(\delta,\chi^2,n) =
\min\left[1,\exp(-\delta-u_B)\right] \ee where \be u_B  =
\frac{\chi^2}{2} + \frac{\chi^4}{4(n+1)} +
\frac{\chi^6}{3(n+1)(n+3)}  + \cdots \ee Note that as the number
of independent samples $n$ gets large, the first term dominates.

The noise level of a system can be characterized by the relative
noise parameter, $f= (\beta \sigma)^2 t/ t_0$, where $t$ is the
computer time spent reducing the noise, and $t_0$ is the computer
time spent on other pursuits, such as optimizing the VMC wave
function or equilibrating the DMC runs.  A small $f$ means little
time is being spent on reducing noise, where a large $f$ means
much time is being spent reducing noise. For a double well
potential, the noise level that gives the maximum
efficiency is around $\beta \sigma \approx 1$, with the optimal
noise level increasing as the relative noise parameter increases \cite{ceperley99}.


We can use multi--level sampling to make CEIMC more efficient
\cite{ceperley95}. An empirical potential is used to
``pre-reject'' moves that would cause particles to overlap and be
rejected anyway. A trial move is proposed and accepted or rejected
based on a classical potential
\be
A_1 = \min \left[1,\frac{T(R\rightarrow R')}{T(R' \rightarrow R)}
      \exp(-\beta \Delta V_{cl})\right]
\ee
where $\Delta V_{cl} = V_{cl}(R') - V_{cl}(R)$ and $T$ is the
sampling probability for a move. If it is accepted at this first
level, the QMC energy difference is computed and accepted with
probability
\be
A_2 = \min \left[1, \exp(-\beta \Delta V_{QMC}
-u_B ) \exp(\beta \Delta V_{cl}) \right]
\ee
where $u_B$ is the noise penalty.

Compared to the cost of evaluating the QMC energy difference,
computing the classical energy difference is much less expensive.
Reducing the number of QMC energy difference evaluations reduces
the overall computer time required. For the molecular hydrogen
system, using the pre--rejection technique with a CEIMC--DMC
simulation results in a first level (classical potential)
acceptance ratio of 0.43, and a second level (quantum potential)
acceptance ratio of 0.52. The penalty method rejects additional
trial moves because of noise. If these rejections are counted as
acceptances (i.e., no penalty method or no noise), then the
second level acceptance ratio would be 0.71.

\section{Energy Differences}

In Monte Carlo it is the energy difference between an old position
and a trial position that is needed. Using correlated sampling
methods it is possible to compute the energy difference with a
smaller statistical error than each individual energy.  We also
need to ensure that that energy difference is unbiased and
normally distributed. In this section we briefly discuss several
methods for computing that difference.

\subsection{Direct Difference}
The most straightforward method for computing the difference in
energy between two systems is to perform independent computations
for the energy of each system. Then the energy difference and
error estimate are given by
\bea
\Delta E &=& E_1 - E_2 \\
\sigma(\Delta E) &=& \sqrt{\sigma_1^2 + \sigma_2^2}
\eea
This method is
simple and robust, but has the drawback that the error is related
to the error in computing a single system. If the nuclear
positions are close together, the energy difference is likely to
be small and difficult to resolve, since $\sigma_1$ and $\sigma_2$
are determined by the entire system.  Hence the computation time
is proportional to the size of the system, not to how far the ions
are moved.

\subsection{Reweighting}
Reweighting is the simplest correlated sampling method. The same
set of sample points, obtained by sampling $p(R)\propto \psi_1^2$
is used for evaluating both energies. The energy difference is
estimated as: \begin{eqnarray}
\label{rew} \nonumber
\Delta E &=& E_1 - E_2  \\ \nonumber
         &=&  \frac{\int dR \ \psi_1^2 \ E_{L1}}{\int dR \ \psi_1^2}-
              \frac{\int dR \ \psi_2^2 \ E_{L2}}{\int dR \ \psi_2^2} \\
\nonumber
         &=&  \frac{\int dR \ p(R) \left(\frac{\psi_1^2}{p(R)}\right) \ E_{L1} }
                   {\int dR \ p(R)\left(\frac{\psi_1^2}{p(R)}\right)}-
           \frac {\int dR\ p(R) \left(\frac{\psi_2^2}{p(R)}\right)\
              E_{L2} }
              {\int dR\ p(R)\ \left(\frac{\psi_2^2}{p(R)}\right). }
\end{eqnarray}
Then an estimate of $\Delta E$ for a finite simulation is
\be
\label{rew_sum}
\Delta E \approx \sum_{R_i \in
\psi^2_1}
     \left[\frac{E_{L1}(R_i)}{N} - \frac{ w(R_i) E_{L2}(R_i)}
    { \sum_i w(R_i)} \right]
\ee
where $w = \psi_2^2/p(R)$.

Reweighting works well when $\psi_1$ and $\psi_2$ are not too
different, and thus have large overlap.  As the overlap between
them decreases, reweighting gets worse due to large fluctuations
in the weights.  Eventually, one or a few large weights will come
to dominate the sum, the variance in $\Delta E$ will be larger
than that of the direct method.  In addition, the distribution of
energy differences will be less normal.

\subsection{Importance Sampling}

In \cite{dewing} we discussed two-sided sampling: the advantages
of sampling the points in a symmetrical way. Here we introduce a
similar method, namely the use of importance sampling to compute
the energy difference. Importance sampling is conceptually similar
to the reweighting described above, however, we optimize the
sampling function $p(r)$ so as to minimize the variance of the
energy difference. If we neglect sequential correlation caused by
the Markov sampling, it is straightforward to determine the
optimal function: \be p^*(R) \propto | \psi_1^2(R) ( E_{L1}(R)-E_1
)- Q \psi_2^2(R) ( E_{L2}(R)-E_2 ) | \label{eq:impsamp} \ee Here
$E_1$ and $E_2$ are the energies of the two systems, and $Q=\int\
\psi_1^2/\int\ \psi_2^2$ is the ratio of the normalization of the
trial functions. In practice, since these are unknown, one
replaces them by a fuzzy estimate of their values, namely we
maximize $p^*(R)$ within an assumed range of values of $E_1 , E_2,
Q$ . A nice feature of the optimal function in (\ref{eq:impsamp})
is that it is symmetric in the two systems leading to correct
estimate of the fixed node energy, even when the nodes for the two
systems do not coincide. Another advantage, is that the
distribution of energy differences is bounded and the resulting
energy difference is guaranteed to be normal. This is because the
sampling probability depends on the local energy. The use of this
distribution with nodes could lead to ergodic problems, but in
practice no such difficulty has been encountered in generating
samples with $p^*$ using ``smart MC'' methods.

As another sampling example, we consider a simplification of the
optimal distribution, namely $P_s(R) \propto  \psi_1^2 + \psi_2^2
$. This is quite closely related to the two-sided method used
earlier \cite{dewing}.  In this distribution, only a single
trajectory is computed, no local energies are needed in the
sampling, and the estimation of the noise is a bit simpler.

\begin{figure}[tb]
\begin{center}
\includegraphics[width=.6\textwidth]{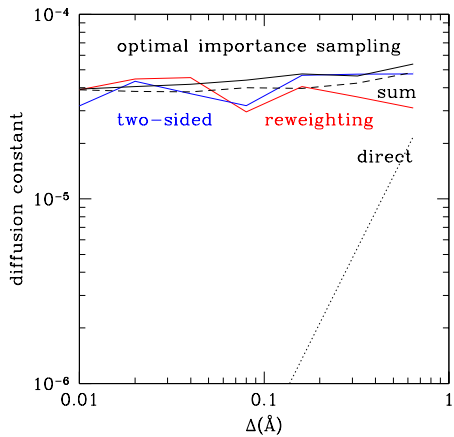}
\end{center}
\caption[]{Efficiency versus importance function on a system with
$N_e=N_p=16$ and $r_s=1.31$. In one system the protons are taken in a sc lattice
and in the other they are displaced randomly, with an average
displacement of $\Delta$.  The diffusion constant is defined as
$\Delta^2/T_{CPU}$ where $T_{CPU}$ is the computer time needed to
calculate the energy difference to an accuracy of $1000~K$.
}\label{dtimes}
\end{figure}

Shown in Figure \ref{dtimes} is the efficiency computed with the
different methods, as a function of the proton step. The curves
show that the various correlated methods have roughly the same
efficiency, which is independent of the size of the proton move.
Correlated methods are more efficient that the direct methods, as
long as the proton are moved less than $\approx 0.8 \AA$. The
optimal importance sampling has about 10\% lower variance than the
reweighting. In addition, the estimates are less biased and
approach a normal distribution much more rapidly\cite{dewing}. We
used the two-sided method and the importance sampling method for
computing energy differences of trial moves with VMC, but only
used the direct method with DMC.

\section{Choice of Trial Wave Function}
An essential part of the CEIMC method is the choice of the trial
wave function. Variational Monte Carlo (VMC) depends crucially on
the trial wave function to find the minimum energy. The trial wave
function is also important in DMC, to reduce the variance and the
projection time, and for accurate nodal surfaces within the
fixed-node method. CEIMC has special demands since optimization of
a trial wave function must be done repeatedly, quickly and without
direct user control.

A typical form of the variational wave function used in QMC is a
Jastrow factor (two body correlations) multiplied by two Slater
determinants of one body orbitals. \be \psi_T =
\exp\left[-\sum_{i<j} u(r_{ij})\right]
{\mathrm{Det}}\left(S^\uparrow\right)
{\mathrm{Det}}\left(S^\downarrow\right) \ee The Slater determinant
is taken from a mean field calculations such as Hartree--Fock or
approximate density functional theory. The cusp condition can be
used to well approximate this at short distances and RPA to
determine the behavior at large distances\cite{foulkes}.

In the molecular phase of hydrogen we estimated that using the orbitals
determined from a separate DFT calculation would have been too slow.
For the molecular phase we resort a simpler alternative namely we used gaussian single
body orbitals, pinned in the center of the molecular bonds.
Optimization of the gaussian, one for each of the molecules,
took much of the computer time. See \cite{dewing00b,dewing} for a detailed
discussion of those results.

For the metallic hydrogen phase, in a previous QMC investigation,
Natoli\cite{natoli93} found that simple plane wave nodes are
inaccurate by 0.05\,eV/atom within the fixed-node approach at the
transition to metallic hydrogen ($r_s=1.31$) necessitating the use
of more accurate (LDA) nodes. However it is inconvenient and
inefficient to solve the LDA equations for each new position of
the ions in CEIMC. In addition, one has to modify the LDA orbitals
to take into account the effect of explicit electron-electron
correlation. The same problem of disordered ionic configurations
arises from zero point motion of the protons in the solid state.
In earlier work on molecular hydrogen, we were unable to use high
quality LDA orbitals when the molecules were angularly
disoriented \cite{natoli95}.

We have recently generalized the backflow and three--body wave
function to a two component system of electrons and protons at
high enough density so that the electrons are delocalized and all
the hydrogen molecules are dissociated. For metallic hydrogen, as
an element without a core, the formalism leading to the improved
wave functions is simplest \cite{chep02}. These wave functions
depend explicitly and continuously on the ionic variables and as a
consequence do not have to be reoptimized for movements of the
ions. These trial functions are a generalization of the backflow
three--body wave functions used very successfully in highly
correlated homogeneous quantum liquids: liquid $^3$He and electron
gas. Backflow trial functions show much improvement over the pair
product getting approximately 80\% of the missing correlation and
even more of the energy when done with the fixed-node method.
Backflow wave functions utilize the power of the QMC sampling
approach: one can calculate properties of such a wave function
without changing the algorithm in an essential way, whileas in
approaches based on explicit integration, one is limited in the
form of the trial function by the ease performing the integration.
We will discuss these backflow functions in more detail below.

\section{Twist Average Boundary Conditions}

Almost all QMC calculations in periodic
boundary conditions have assumed that the phase of the wave
function returns to the same value if a particle goes around the
periodic boundaries and returns to its original position. However,
with these boundary conditions, delocalized fermion systems
converge slowly to the thermodynamic limit because of shell
effects in the filling of single particle states. One can allow
particles to pick up a phase when they wrap around the periodic
boundaries, \be \Psi(\vec{r}_1 + L  \hat{\vec{x}}, \vec{r}_2,
\cdots )= \E^{i\theta_x} \Psi(\vec{r}_1,\vec{r}_2, \cdots ).
\label{TBC} \ee The boundary condition $\theta = 0$ is periodic
boundary conditions (PBC), and the general condition with $\theta
\neq 0$, twisted boundary conditions (TBC). The use of twisted
boundary conditions is commonplace for the solution of the band
structure problem for a periodic solid, particularly for metals.
In order to calculate properties of an infinite periodic solid,
properties must be averaged by integrating over the first
Brillouin zone.

For a degenerate Fermi liquid, finite-size shell effects are much
reduced if the twist angle is averaged over: twist averaged
boundary conditions (TABC). This is particularly important in
computing properties that are sensitive to the single particle
energies such as the kinetic energy and the magnetic
susceptibility. By reducing shell effects, much more accurate
estimations of the thermodynamic limit for these properties can be
obtained. What makes this even more important is that the most
accurate quantum methods have computational demands which increase
rapidly with the number of fermions. Examples of such methods are
exact diagonalization  (exponential increase in CPU time with N),
variational Monte Carlo (VMC) with wave functions having backflow
and three-body terms \cite{kwon2d,kwon3d} (increases as $N^4$),
and transient-estimate and released-node Diffusion Monte Carlo
methods \cite{cep84} (exponential increase with N). Methods which
can extrapolate more rapidly to the thermodynamic limit are
crucial in obtaining high accuracy.

Twist averaging is especially advantageous in combination with
CEIMC (i.e.  QMC) because the averaging does not necessarily slow
down the evaluation of energy differences, except for the
necessity of doing complex rather than real arithmetic. In a
metallic system, such as hydrogen at even higher pressure when it
becomes a simple metal, results in the thermodynamic limit require
careful integration near the Fermi surface because the occupation
of states becomes discontinuous. Within LDA this requires
``k--point'' integration, which slows down the calculation
linearly in the number of k-points required. Within QMC such
k-point integration takes the form of an average over the (phase)
twist of the boundary condition and can be done in parallel with
the average over electronic configurations without significantly
adding to the computational effort.  We typically spawn about 100
distinct QMC processes, run for a fixed time, and then average the
resulting properties.

\section{Fluid Molecular Hydrogen}

We now describe our calculations on  liquid molecular hydrogen.
First of all, we examine the accuracy of several methods for
computing total energy. We took several configurations from PIMC
simulations at 5000\,K at two densities ($r_s = 1.86$ and $r_s =
2.0$), and compared the electronic energy using VMC, DMC, DFT-LDA,
and some empirical potentials. The DFT--LDA results  were obtained
from a plane wave code using an energy cutoff of 60\,Rydbergs, and
using the $\Gamma$ point approximation \cite{ogitsu00}. The
empirical potentials we used are the
Silvera--Goldman \cite{silvera78} and the
Diep--Johnson \cite{diep00a,diep00b}. To these we added the energy
from the Kolos \cite{kolos64} intramolecular potential to get the
energy as a function of the bond length variations. The
Silvera--Goldman potential was obtained by fitting to low
temperature experimental data, with pressures up to 20\,Kbar, and
is isotropic. The Diep--Johnson potential is the most recent in a
number of potentials for the isolated H$_2$--H$_2$ system. It was
fit to the results of accurate quantum chemistry calculations for
a number of H$_2$--H$_2$ configurations and included anisotropic
effects in the potential.

\begin{figure}[t]
\begin{center}
\includegraphics[width=.6\textwidth]{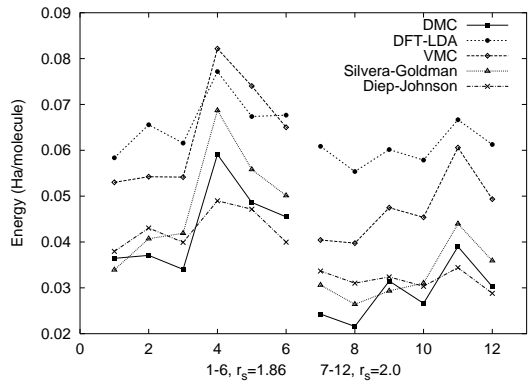}
\end{center}
\caption{ Electronic energy for several configurations computed by
various methods. The energy is relative to an isolated H$_2$
molecule.
 \label{fig:pimc-cfgs}}
\end{figure}

The energies relative to an isolated H$_2$ molecule are shown in
Figure \ref{fig:pimc-cfgs}.  The first thing we notice is that the
classical potentials are more accurate than VMC or DFT. The
Silvera--Goldman does a good job of reproducing the DMC results.
Some of the failures of the SG potential can be attributed to the
lack of anisotropy. The isolated H$_2$--H$_2$ potential
(Diep--Johnson) has much weaker interactions, compared with
interactions in a denser system.

The PIMC method itself gives an average energy of about $0.07(3)$\,Ha
for both densities. Improvements in the fermion nodes and in
other aspects of the PIMC calculation appear to lower the
energy \cite{militzer00a,militzer00b,militzer-thesis00}, although
the error bars are still quite large.  The PIMC energy is in rough
agreement with the DMC energy.

As mentioned above, we used the Silvera--Goldman potential for
pre--rejection. As seen in the Figure \ref{fig:pimc-cfgs}, it
resembles the DMC potential even though it lacks anisotropy. Each
trial move consisted of moving multiple molecules (usually four).
This increased efficiency by amortizing the cost of the QMC
calculation. Each molecular move was decomposed into a translation
of the center of mass, a rotation of the molecule, and a change in
the bond length.

Shown in Tables 1--2 are CEIMC results at three state points, two
of which can be compared with the gas gun data of Holmes {\it et
al}. \cite{holmes95}. The pressure is given in table
\ref{gasgundata} with results from the gas gun experiments, the
free energy model of Saumon and Chabrier
\cite{saumon91,saumon92,saumon95}, from simulations using the
Silvera--Goldman potential, and from our CEIMC simulations. These
state points are in the fluid molecular H$_2$ phase. For the gas
gun experiments, the uncertainties in the measured temperatures
are around 100--200\,K. The experimental uncertainties in the
volume and pressure were not given, but previous work indicates
that they are about 1--2\% \cite{nellis83}.

We did CEIMC calculations using VMC or DMC for computing the
underlying electronic energy, which are the first such QMC
calculations in this range. The simulations at $r_s=2.1$ and
$r_s=1.8$ were done with 32 molecules, and the simulations at
$r_s=2.202$ were done with 16 molecules. We see that the pressures
from VMC and DMC are very similar, and that for $r_s=2.1$ we get
good agreement with experiment.

There is a larger discrepancy with experiment at $r_s=2.202$. The
finite size effects are fairly large, especially with DMC. We also
did simulations at $r_s=2.1$ with 16 molecules and obtained
pressures of $0.264(3)$\,Mbar for CEIMC--VMC and $0.129(4)$\,Mbar for
CEIMC--DMC. The Silvera--Goldman potential showed much smaller
finite size effects than the CEIMC simulations, so we conclude
that the electronic part of the simulation is largely responsible
for the observed finite size effects.

\begin{figure}[t]
\begin{tabular}{cc}
 \includegraphics[width=.6\textwidth]{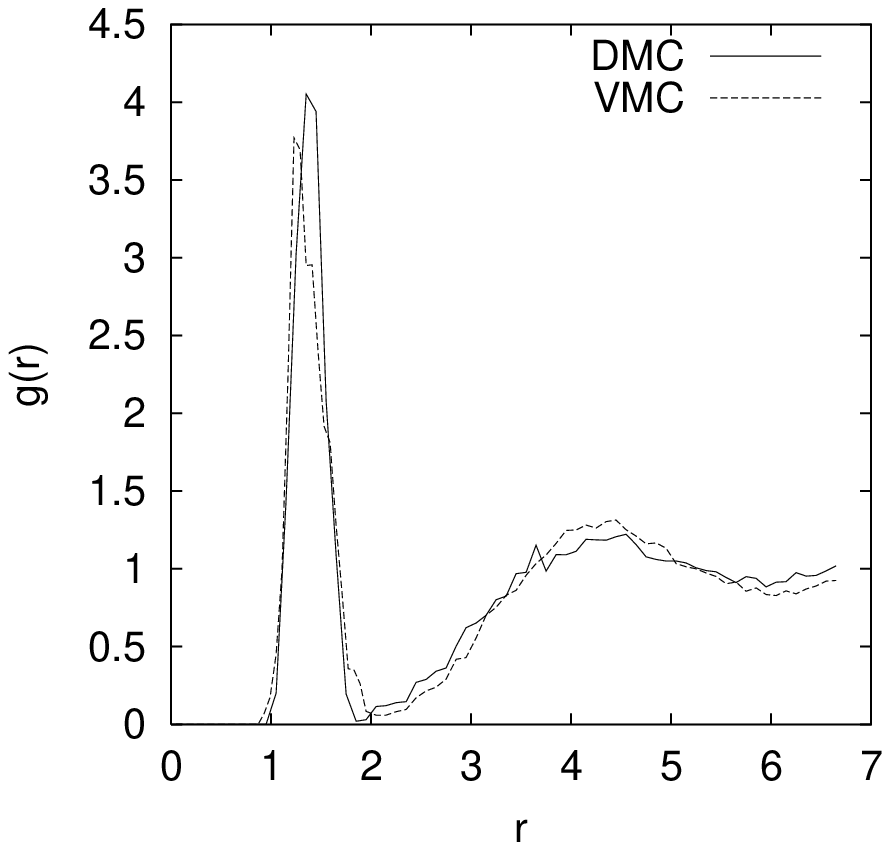}
 &\includegraphics[width=.6\textwidth]{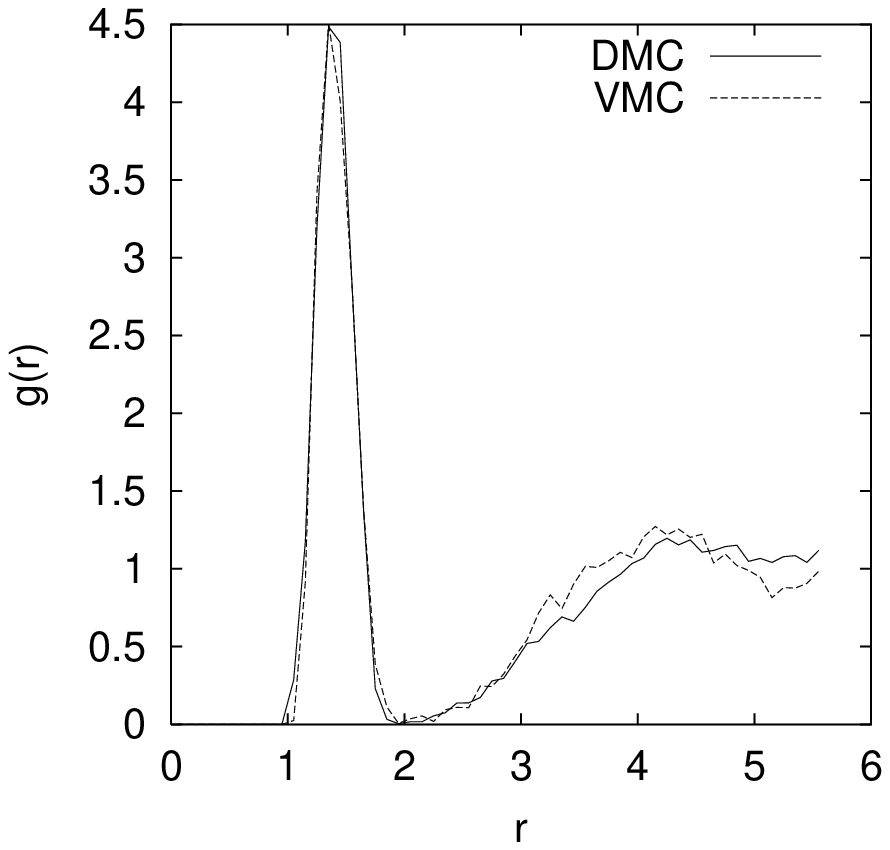}\\
(a) & (b) \\
\end{tabular}
\caption{Proton pair distribution, $g(r)$, for {\bf(a)} $r_s = 2.1$ and
T=4530\,K {\bf(b)} $r_s = 2.202$ and T=2820\,K
 \label{fig:gr-results}}
\end{figure}

The energies for all these systems are given in Table
\ref{h_energy}. The energy at $r_s=2.1$ with 16 molecules for
CEIMC--VMC is $0.0711(4)$\,Ha and for CEIMC--DMC is $0.0721(8)$\,Ha.
The proton--proton distribution functions
comparing CEIMC--VMC and CEIMC--DMC are shown in Figure
\ref{fig:gr-results}. The VMC and DMC distribution functions look
similar, with the first large intramolecular peak around $r=1.4$
and the intermolecular peak around $r=4.5$.

\begin{table}
\begin{center}
\caption{Pressure from simulations and shock wave experiments
 \label{gasgundata}}
 \begin{tabular}{|cc|ccccc|} \hline
r$_s$  &       T(K)   & \multicolumn{5}{c|}{Pressure (Mbar)} \\
     &      &   Gasgun  &  S--C &  S--G & CEIMC--VMC & CEIMC--DMC \\\hline
 2.100        &   4530    &    0.234
    &  0.213  & 0.201  & 0.226(4)  & 0.225(3)  \\
 2.202       &   2820    &    0.120
    & 0.125   &  0.116      & 0.105(6)   &  0.10(5)  \\
 1.800      & 3000 & - & - & 0.528 & - & 0.357(8)  \\ \hline
 \end{tabular}
\end{center}
\end{table}

\begin{table}
\begin{center}
\caption{Energy from simulations and models,
 relative to the ground state of an
isolated H$_2$ molecule. The H$_2$ column is a single thermally
excited molecule plus the quantum vibrational KE.
 \label{h_energy}}
 \begin{tabular}{|cc|ccccc|}
\hline
r$_s$     &    T(K)   & \multicolumn{5}{c|}{Energy (Ha/molecule)} \\
     &         &   H$_2$  &  S--C &  S--G & CEIMC--VMC & CEIM-C-DMC \\
\hline
 2.100      &   4530    &    0.0493
    &  0.0643  & 0.0689  & 0.0663(8)  & 0.0617(2)  \\
 2.202       &   2820    &    0.0290
    & 0.0367   &  0.0408      & 0.0305(8)   &  0.0334(9)  \\
 1.800      & 3000 & 0.0311   & - & 0.0722   & - & 0.048(1)\\
\hline
 \end{tabular}
\end{center}
\end{table}

The CEIMC--VMC simulations at $r_s=1.8$ and 3000\,K never
converged. Starting from a liquid state, the energy decreased
during the entire simulation. Visualization of the configurations
revealed that they were forming a plane. It is not clear whether
it was trying to freeze, or forming structures similar to those
found in DFT--LDA calculations with insufficient Brillouin zone
sampling\cite{hohl93,kohanoff97}. (note that the molecular
hydrogen calculations were done at the $\Gamma$ point.)The
CEIMC--DMC simulations did not appear to have any difficulty, so
it seems the VMC behavior was due to inadequacies of the wave
function.

\begin{figure}[t]
\begin{center}
\includegraphics[width=.6\textwidth]{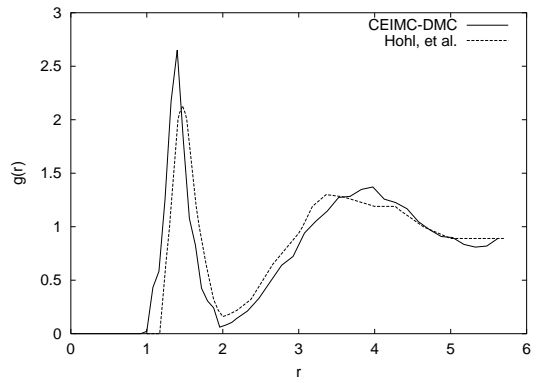}
\end{center}
\caption{ The proton pair distribution function, $g(r)$, near $r_s
= 1.8$ and T=3000K.
 \label{fig:rs18}}
\end{figure}

Hohl {\it et al.}\cite{hohl93} did DFT--LDA simulations at
$r_s=1.78$ and T=3000\,K, which is very close to our simulations
at $r_s=1.8$.  The resulting proton-proton distribution functions
are compared in Figure \ref{fig:rs18}. The CEIMC distribution has
more molecules and they are more tightly bound. The discrepancy
between CEIMC and LDA in the intramolecular portion of the curve
has several possible causes. On the CEIMC side, it may be due to
lack of convergence or to the molecular nature of the wave
function, which does not allow dissociation. The shift of the
molecular bond length peak can be accounted for because LDA is
known to overestimate the bond length of a free hydrogen
molecule \cite{hohl93} which would account for the shifted location
of the bond length peak. The deficiencies of LDA may account for
it preferring fewer and less tightly bound molecules.

\section{The atomic-metallic phase}

In this section, we describe preliminary results for metallic
atomic hydrogen from a recent implementation of the method using
an improved wave functions including threebody and backflow terms
and taking advantage of averaging over the twist angle to minimize
size effects.

\subsection{Trial wave function and optimization}

We have seen that an important part of the CPU time is needed in
the optimization of the molecular trial wave functions which
contain a number of variational parameters proportional to the
number of molecules and which need to be optimized individually
for each protonic steps. A major improvement in the efficiency of
the method can be achieved by using more sophisticated wave
functions, namely analytic functions in terms of the proton
positions which move with the protons and which depend on few
variational parameters (about 10, regardless of the number of
particles). Moreover, one can explore the possibility of
optimizing the variational parameters only once at the beginning
of the calculation, either on ordered or disordered protonic
configurations, and using the optimized wave function during the
simulation. In this section, we consider hydrogen at densities at
which molecules are dissociated ($r_s \le 1.31$) and the system is
metallic. We will therefore avoid the complications arising from
the presence of bound states (either molecular or atomic). In this
case one can show that improved wave functions with respect to the
simple Slater-Jastrow form includes backflow and threebody terms
between electrons and protons \cite{chep02} in a very similar
fashion as the ones used by Kwon et al.\cite{kwon2d,kwon3d} for
the electron gas. We assume a trial wave function of the form \be
\label{eq:psit} \Psi_T(\vec{R})=det(e^{i{\kvec_i\cdot\xvec_j}})
exp\left(-\sum_{i<j}^N
\tilde{u}(r_{ij})-\frac{\lambda_T}{2}\sum_{l=1}^N{\bf
G}(l)\cdot{\bf G}(l)\right) \ee where \bea
\xvec_i=\rvec_i+\sum_{j\ne i}^N \eta(r_{ij})(\rvec_i-\rvec_j) \\
{\bf G}(l)=\sum_{i\ne l}^N \xi(r_{li})(\rvec_l-\rvec_i) \\
\tilde{u}(r)=u(r)-\lambda_T \xi^2(r) r^2
\eea
with
\bea
\eta(r)=\lambda_b exp[-(r-r_b)^2/w_b^2]\\
\xi(r)=exp[-(r-r_T)^2/w_T^2]
\eea and $u(r)$ is an optimized version of the RPA pseudopotential \cite{ca87}.

In what follows we only consider the effect of electron--proton
backflow and electron--proton--proton three body terms, while the
electronic part of the wave function is of the simple
Slater--Jastrow form. To establish the goodness of this wave
function for metallic hydrogen we perform VMC and DMC calculation
for 16 protons on a bcc lattice at $r_s=1.31$. In table [3] 
we compare the energy and the variance of the
local energy of this wave function with data obtained with the
simple Slater-Jastrow wavefunction and with an improved
wavefunction in which a determinant of single body orbital from a
separate LDA calculation has been used \cite{natoli93}. From these
results we infer that the nodes of the new wavefunction are as
accurate as the LDA nodes.

\begin{table}
\label{tab:rs1.31}
\caption{$r_s=1.31$. Energy and variance for 16 protons in the bcc lattice. SJ, SJ3B and LDA
indicate optimized Slater--Jastrow, optimized Slater--Jastrow+three--body+backflow,
and LDA nodes respectively. Energies per particle are in Rydbergs.}
\begin{center}
\renewcommand{\arraystretch}{1.4}
\setlength\tabcolsep{5pt}
\begin{tabular}{|c|ccc|}
\hline\noalign{\smallskip}
  $$ & $E_{VMC}$  & $\sigma_{VMC}^2$& $E_{DMC}$ \\
\hline
 SJ  & -0.4754(2) &   0.0764(9)     & -0.4857(1) \\
 SJ3B& -0.4857(2) &   0.0274(2)     & -0.4900(1) \\
 LDA & -0.4870(10)&                 & -0.4890(5) \\
\hline
\end{tabular}
\end{center}
\end{table}

Having established that our wavefunction at $r_s=1.31$ is as good
as the most accurate wavefunction used for metallic hydrogen so
far, we continue our study at slightly higher density, namely
$r_s=1$. It can be shown that the above form of the wavefunction
is obtained using perturbation theory from the high density limit
and we expect that its accuracy improves for decreasing $r_s$.

\begin{figure}[]
\begin{center}
\includegraphics[width=.5\textwidth]{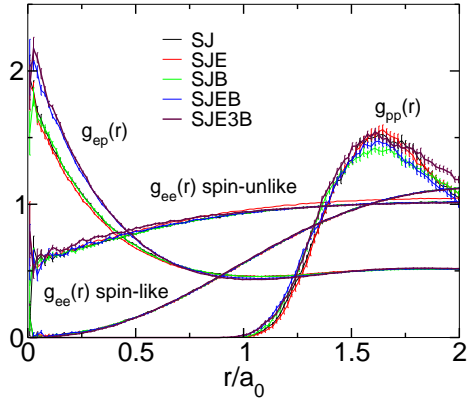}
\end{center}
\caption[]{ $r_s=1, T=5000\,K, N_e=N_p=16$ spin unpolarized. Pair correlations functions with various
trial wave functions. The entries in the legend corresponds to the entries in table \ref{tab:rs1t5n32}.
The $g_{ep}(r)$ have been shifted downward by 0.5 for sake of clarity.}
\label{fig:rs1t5n32gr}
\end{figure}

We first perform a number of optimizations of the trial wave
function. Beside the RPA e--p Jastrow, we consider an extra 2 body
(e--p) gaussian term with two variational parameters ($\lambda_e$
and $w_e$). In table \ref{tab:rs1t5n32} we report the values of
the variational parameters obtained minimizing a linear
combination of the local energy and its variance over a set of
different protons and electrons configurations. Typically 1000
configurations have been used, by saving a configuration after  10
or 20 protonic steps during a previous run. We also studied the
relative importance of the different terms in the trial
wavefunction by performing calculations with partially improved
wave functions. In Figure \ref{fig:rs1t5n32gr} we compare the pair
correlation functions for the various calculations in table
\ref{tab:rs1t5n32}. No significant difference is observed in the
electron--electron and in the proton--proton pair correlation
functions among different forms of the trial function. We can see
that the cooperative effects of the optimized Jastrow factor and
the backflow term are responsible of an enhancing of
electron-proton correlation as seen in the $g_{ep}(r)$. Inclusion
of three--body terms lowers the energy but does not change the
pair correlations.

\begin{table}
\caption{$r_s=1, T=5000\,K, N_p=N_e=16$. Optimized values of the
variational parameters for the VMC trial function. The values are
obtained minimizing local energy and variance for 1000 different
equilibrium configurations.} \label{tab:rs1t5n32}
\begin{tabular}{|c|ccccccccc|}
\hline
    & $\lambda_b$& $r_b$   & $w_b$  &$\lambda_T$&$r_T$&$w_T$&$\lambda_e$& $w_e$& E(a.u.) \\
\hline
SJ  &    --     &   --    &  --    &  ---  &  --    & --   & --     &  --    & -0.117(1) \\
SJE &   ---     &   ---   &  ---   &  --   &  --    &  --  & 0.06167& 0.9497 & -0.1180(4) \\
SJB & -0.60824  & -1.3726& 1.44822 &  --   &  --    &  --  &    --  &  --    & -0.1207(4) \\
SJEB& -0.45828  & -0.60202& 0.91273&  --   &  --    &  --  & -0.0874& 1.7002 & -0.1227(5) \\
SJE3B& -0.4671   & -0.6217 & 1.0193 &-2.4676&-1.0917 &3.0029&-0.0844  &1.5130 & -0.1238(2)\\
\hline
\end{tabular}
\end{table}

In principle, the optimization study could be repeated at each
temperature needed for the CEIMC simulation. It is therefore of
practical interest to investigate the transferability at finite
temperature of wavefunctions optimized for the lattice
configurations of protons.

An additional ingredient discussed above and crucial for a
metallic system are finite size effects. It has been shown
recently \cite{lzc01} that the very irregular behavior of the
energy versus N observed in the presence of a Fermi surface can be
reduced to the classical $1/N$ behaviour by averaging over the
twist of the wavefunction. We have implemented the twist averaged
boundary conditions in the calculation of the energy differences
needed to make the protonic moves in the CEIMC. We average over
1000 different twist angles in the three dimensional interval
$(-\pi,\pi)$ found to be sufficient in the electron
gas\cite{lzc01}. The additional issue of whether optimization of
the variational parameters need to be done with or without twist
averaging was investigated. We compare in table
\ref{tab:sph_vs_tas_vf_T} the results of single phase optimization
and phase averaging optimizations for protons in the bcc lattice
and at T=5000\,K. The fourth row is the result of an optimization
with twist averaging at T=0\,K, while the fifth row is a run with
the values of the variational parameters optimal for the $\Gamma$
point, always at T=0\,K. We observe an excellent agreement of the
energies and we conclude that we can safely optimize the wave
function using the $\Gamma$ point and use the obtained variational
parameters for all twist angles.

The sixth row in the table is the result of a simulation at
T=5000\,K using the values of the variational parameters optimal
for T=0\,K (bcc lattice). The energy should be compared with the
result of the entry SJE3B in table \ref{tab:rs1t5n32}. The
difference in energy is within error bars and indicates that we
can safely optimize the wave function on lattice configurations
for use at finite temperature to avoid repeating the optimization
at each temperature. In the last two rows of table
\ref{tab:sph_vs_tas_vf_T} we report results of two runs at
T=5000\,K with twist averaging. In the first run the variational
parameters optimized in the bcc configuration and with the
$\Gamma$ point have been used. In the second one new values of the
parameters, obtained by optimization over a set of configurations
stored in the previous twist-averaged run, have been used. The
excellent agreement on the energy (and on the variance of the
local energy, not shown in the table) confirms that optimization
of the variational parameters can be safely performed in the
lattice configuration and with a single phase.

\begin{table}
\caption{$r_s=1, N_p=N_e=16$ spin unpolarized.
Optimized values of the variational parameters for the VMC trial function.
The values are obtained minimizing local energy and variance for 1000 different
equilibrium configurations.}
\label{tab:sph_vs_tas_vf_T}
\begin{tabular}{|cc|ccccccccc|}
\hline\noalign{\smallskip}
T(K)&\#phases&$\lambda_b$&$r_b$&$w_b$&$\lambda_3$&$r_3$&$w_3$&$\lambda_e$&$w_e$&E(a.u.) \\
\noalign{\smallskip}
\hline
\noalign{\smallskip}
  0 &    1  &  --   &   --  &  --  &  ---  &  --   & --   & --    &  --  & -0.1306(2) \\
  0 &    1  &-0.2574&-0.2172&0.7623&-2.3742&-1.8150&1.9694&-0.0496&1.7937& -0.1353(1) \\
\noalign{\smallskip}
\hline
\noalign{\smallskip}
  0 &  1000 &  --   &  --   &  --  &  --   &  --   &  --  &   --  &  --  & -0.1779(1) \\
  0 &  1000 &-0.2386&-0.1757&0.6613&-2.2609&-1.8326&3.3130&-0.0475&2.0337& -0.18254(3) \\
  0 &  1000 &-0.2574&-0.2172&0.7623&-2.3742&-1.8150&1.9694&-0.0496&1.7937& -0.18253(3) \\
\noalign{\smallskip}
\hline
\noalign{\smallskip}
5000&    1  &-0.2574&-0.2172&0.7623&-2.3742&-1.8150&1.9694&-0.0496&1.7937& -0.1237(2) \\
5000&  1000 &-0.2574&-0.2172&0.7623&-2.3742&-1.8150&1.9694&-0.0496&1.7937& -0.1708(3) \\
5000&  1000 &-0.4611&-0.7339&1.1287&-2.0402&-2.2098&3.0213&-0.0949&1.2389& -0.1709(3) \\
\noalign{\smallskip}
\hline
\noalign{\smallskip}
\end{tabular}
\end{table}

\subsection{Comparison with PIMC}

In order to establish the accuracy of the CEIMC method, we compare
CEIMC and PIMC results at high temperatures and pressures. To
eliminate the ``fermion sign problem'', the R-PIMC technique for
fermions assumes the nodal surfaces of a trial density matrix. In
most of the applications, free particle nodal surfaces, either
temperature dependent or in the ground state, have been
used\cite{helium3-92,pcbm94,como95}. More recently, variational
nodes which account for bound states have been implemented in the
study of the plasma phase transition
\cite{militzer00a,militzer00b}. However, the use of temperature
dependent nodes, which break the imaginary time translational
symmetry, is limited to quite high temperature, $T \ge 0.1 T_F$
where $T_F(a.u.)=1.84158/r_s^2$ is the electronic Fermi
temperature. Below this threshold, the Monte Carlo sampling
becomes extremely inefficient and the method impractical. This
pathology is not encountered when using ground state nodes, which
preserve the original imaginary-time symmetry and are expected to
become as accurate as the temperature dependent nodes at low
enough temperature. At $r_s=1$ ($0.1 T_F=0.18158 a.u.\approx$
57300\,K), we perform calculation at T=10000\,K and T=5000\,K and
we exploit the PIMC with free particle ground state nodes. In
table \ref{tab:bomcVSpimc} we compare energies and pressure from
PIMC and CEIMC simulations. At T=10000\,K, two different PIMC
studies are reported, with $M=500$ and $M=1000$ time slices
respectively, which correspond to $\tau=0.063(a.u.)^{-1}$ and
$\tau=0.0315(a.u.)^{-1}$. The smaller value satisfy the empirical
criteria for good convergence $\tau\le0.05/r_s^2 (a.u.)^{-1}$ we
have established in the plasma phase at higher temperature
\cite{pcbm94}. At T=5000\,K only $M=1000$ has been used and
therefore the convergence with the number of time slices is
limited.

We see small differences between PIMC and CEIMC. In particular,
the electronic kinetic energy in PIMC is always slightly higher
than in CEIMC. At the same time, CEIMC determined potential energy
is lower than the PIMC value and this results in a significantly
lower total energy of CEIMC compared to PIMC. The difference
between PIMC and CEIMC seems to decrease with temperature. To
judge the quality of these results, we should keep in mind
advantages and limitations of each method. PIMC uses the ``exact''
bosonic action, the electrons are at finite temperature and
excited states are taken into account, although in a approximate
way because of the simplified nodal restriction: its approximation
for the nodal surface is a Slater determinant of plane waves.
CEIMC instead assumes a trial functions which, at the correlation
(bosonic) level is certainly an approximation to the true bosonic
action used in PIMC. Moreover, the electrons are in their ground
state by construction. However, the trial wavefunction in CEIMC is
better (for the ground state) than the one used in PIMC. Because
of these differences we think that the comparison between the two
methods shows agreement although a more detailed investigation is
in order.

\begin{table}
\caption{$r_s=1, N_p=N_e=18$. Comparison between PIMC and CEIMC
methods at T=10000\,K and T=5000\,K.}
\renewcommand{\arraystretch}{1.4}
\begin{center}
\begin{tabular}{|c|ccccccc|}
\hline\noalign{\smallskip}
method&$T/10^3(K)$& M &$K_e/N_e$&  $K_t/N$ & $V_N/N$   &   $E/N$   &   $P$ \\
\noalign{\smallskip}
\hline
\noalign{\smallskip}
 PIMC &10&500&1.477(9)& 0.763(5)  & -0.7771(6) & -0.0141(6)  & 0.119(2)\\
 PIMC &10&1000&1.48(1)& 0.764(6)  & -0.7820(8) & -0.0180(7)  & 0.119(2) \\
 CEIMC&10& -- &1.3767(4)&0.7121(2)& -0.7995(2) & -0.0874(4) & 0.0994(1) \\
\noalign{\smallskip}
\hline
\noalign{\smallskip}
 PIMC & 5&1000&1.39(1)~~& 0.707(5)  & -0.791(1)  & -0.084(6)  & 0.099(2) \\
 CEIMC& 5& -- &1.317(1) &0.6703(3)  & -0.7939(2) & -0.1236(2) & 0.0870(8)\\
\noalign{\smallskip}
\hline
\noalign{\smallskip}
\end{tabular}
\end{center}
\label{tab:bomcVSpimc}
\end{table}

\begin{figure}[]
\begin{center}
\includegraphics[width=.5\textwidth]{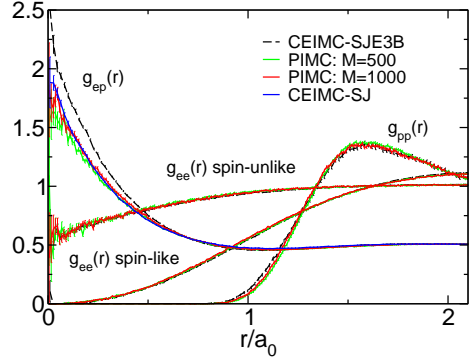}
\end{center}
\caption[]{Pair correlation functions at T=10000\,K. Comparison between PIMC and CEIMC.
$g_{ep}(r)$ have been shifted by $-0.5$ for sake of clarity.}
\label{fig:bomcVSpimc10}
\end{figure}

\begin{figure}[]
\begin{center}
\includegraphics[width=.5\textwidth]{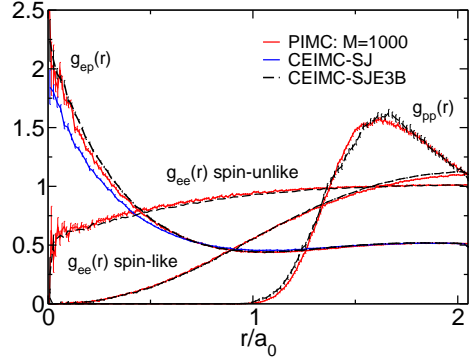}
\end{center}
\caption[]{Pair correlation functions at T=5000\,K. Comparison between PIMC and CEIMC.
$g_{ep}(r)$ have been shifted by $-0.5$ for sake of clarity.}
\label{fig:bomcVSpimc5}
\end{figure}

Comparison between PIMC and CEIMC for the pair correlation
functions at T=10000\,K and T=5000\,K is given in Figures
\ref{fig:bomcVSpimc10} and \ref{fig:bomcVSpimc5} respectively. In
Figure \ref{fig:bomcVSpimc10} we first note a good agreement
between the two PIMC calculations which show that pair correlation
functions are much less sensible to finite imaginary time step
errors. In general for all correlation functions except the
electron-proton ones, the agreement between PIMC and CEIMC is
excellent.

At T=10000\,K the electron--proton pair correlation function from
PIMC is in very good agreement with the result of the CEIMC method
where a simple Slater--Jastrow trial wave function is used.
Improving the trial wave function as discussed in the previous
subsection worsen the agreement. The opposite behavior is observed
at T=5000\,K where the better agreement between PIMC and CEIMC is
observed with the improved wave function (SJE3B in the Figure). We
interpret this behaviour as follows: at lower temperature the
improved trial wave function (\ref{eq:psit}) provides the
``correct'' electron--proton correlation (through the combined
effect of the optimized Jastrow and the electron--proton backflow,
see the discussion relative to Figure \ref{fig:rs1t5n32gr} in the
previous subsection). At this temperature the electronic thermal
effects on the electron--proton correlation are quite small and
the electronic ground state as provided by CEIMC is quite
accurate. Instead at higher temperatures electron--proton
scattering is influenced by excited electronic states which are
not considered in the CEIMC method. As a result the
electron--proton pair correlation function shows a weaker
correlation near the origin and it is in better agreement with the
CEIMC result with the Slater--Jastrow trial function rather than
with the CEIMC result for the improved trial function.

\subsection{Hydrogen equation of state and solid--liquid phase transition of the protons}

We present in this subsection, results for the equation of state
(EOS) of hydrogen in the metallic phase including the
solid--liquid transition of protons. These results are preliminary
in various respects. Firstly, the electrons are treated at the
variational level and no use of Projection/Diffusion Monte Carlo
was attempted. Secondly, the protons are considered as classical
point particles although it is well known that zero point motion
at such high pressure can be significant (at least around the
phase transition). Finally, we have data at a single density
($r_s=1$) and for a single system size, namely $N_e=N_p=32$
(compatible with the fcc lattice) and we cannot address, at this
stage, the issue of the relative stability of different crystal
structures. Nonetheless, we believe these results are interesting
because they show the applicability and provide a benchmark of the
method.

In table \ref{tab:rs1n64details}, we report various details of the
simulations such as the maximum amplitude of the protonic step in
units of the Bohr radius $\Delta_p$, the total number of
electronic steps per protonic step $M_{el}$, the relative noise
level for the penalty method $(\beta\sigma)^2$, the acceptance for
the protonic move and the noise rejection ratio $\eta$
\cite{dewing}. A measure of the computational efficiency can be
defined as the proton diffusion in configurational space with
respect to the CPU time $D_p=(\sum_p [\Delta R_p]^2)/T_{CPU}$. In
the table we report the values obtained in our simulations in
units of $a_0^2/sec$. In addition, the CPU time per protonic step,
the number of processors and the machine used are
reported\footnote{Beowulf is a IBMx330 cluster with
PentiumIII/1.13GHz in CINECA--ITALY
(www.cineca.it/HPSystems/Resources/LinuxCluster), origin3800 is a
SGI--origin3800 with R14000/500MHz in CINES--FRANCE (www.cines.fr)
and platinum is a IBMx330 cluster with PentiumIII/1GHz in
NCSA--USA
(www.ncsa.uiuc.edu\-/UserInfo\-/Resources\-/Hardware\-/IA32LinuxCluster\-/TechSummary).}.

Note that $M_{el}=15000$ is the minimum number of electronic steps
needed to average over 1000 different twist angles. Except at the
lowest temperatures ($T\le$500\,K) such large number of electronic
steps would not be necessary in order to reduce the noise level.
Further improvements in efficiency could be gained by reducing the
number of electronic steps or the number of angles averaged over
for $T > 500\,K$.

\begin{table}
\caption{Simulation details: $r_s=1, N_p=N_e=32$.  $\Delta_p$ is
the maximum amplitude of the protonic step in units of the Bohr
radius, $M_{el}$ is the total number of electronic steps per
protonic step, $(\beta\sigma)^2$ is the relative noise level
entering in the penalty method, $P_{acc}$ is the average
acceptance of the protonic moves, $\eta$ is the noise rejection
ratio defined earlier, $D_p$ is the diffusion constant in protonic
configurational space with respect to CPU time.}
\renewcommand{\arraystretch}{1.4}
\begin{center}
\begin{tabular}{|cccccccccc|}
\hline\noalign{\smallskip}
$T (K)$&$\Delta_p$&$M_{el}$&$(\beta\sigma)^2$&$P_{acc}$&$\eta$&$D_p\times 10^{4}$& time/step(sec)& \#proc & machine\\
\noalign{\smallskip}
\hline
\noalign{\smallskip}
 5000 & 0.03 & 15000 &0.037(4)&0.80&0.0084&1.9(2)&5.96&32&beowulf \\
 4000 & 0.03 & 15000 &0.092(8)&0.77&0.013 &3.8(3) &5.93&32&beowulf \\
 3000 & 0.025& 15000 &0.10(2) &0.76&0.012 &3.2(3) &10.3&16&origin3800 \\
 2000 & 0.03 & 15000 &0.29(5) &0.68&0.033 &2.6(3) &10.3&16&origin3800 \\
 1000 & 0.02 & 15000 &0.30(3) &0.64&0.06  &3.6(3) &10.3&16&origin3800 \\
  700 & 0.02 & 15000 &0.418(4)&0.55&0.10  &3.6(3) &9.90&16&origin3800 \\
  500 & 0.02 & 15000 &0.747(5)&0.43&0.16  &0.47(8)&8.93&32&platinum \\
  300 & 0.015& 18000 &0.855(9)&0.39&0.21  &0.18(1)&7.02&32&platinum \\
\noalign{\smallskip}
\hline
\noalign{\smallskip}
\end{tabular}
\end{center}
\label{tab:rs1n64details}
\end{table}

\begin{figure}[b]
\begin{center}
\includegraphics[width=.5\textwidth]{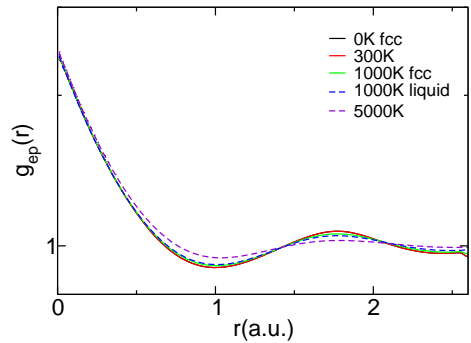}
\end{center}
\caption[]{$r_s=1, N_e=N_p=32$ spin unpolarized. Temperature
dependence of the electron--proton pair correlation functions. The
y scale is logarithmic.} \label{fig:rs1n64grep}
\end{figure}

\begin{figure}[]
\begin{center}
\includegraphics[width=.5\textwidth]{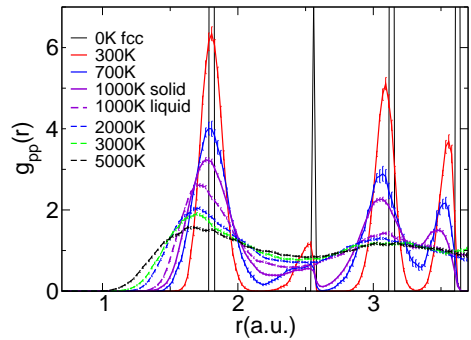}
\end{center}
\caption[]{$r_s=1, N_e=N_p=32$ spin unpolarized. Temperature dependence of the proton--proton
pair correlation functions. The difference between the crystal and the liquid is clearly seen.}
\label{fig:rs1n64grpp}
\end{figure}

In table \ref{tab:rs1n64} we report thermodynamic quantities for
the system at various temperatures in the range
$T\in$[300\,K,5000\,K]. The corresponding electron--protons and
protons--protons pair correlation functions are given in Figures
\ref{fig:rs1n64grep} and \ref{fig:rs1n64grpp} respectively. At
each temperature, equilibrium runs of at least 20000 protonic
steps have been performed. Statistics are collected every 20--50
steps. Besides energies and pressure we compute the Lindemann
ratio $\gamma$ for the fcc structure. In the upper part of the
table, i.e. at higher temperature, we report in the last column
the status of the system. At T=1500\,K the system initially in the
fcc configuration is found to melt after few thousands steps.
Conversely at T=500\,K we started the simulation in a disordered
state and the system spontaneously ordered into the fcc structure.
At T=1000\,K instead, a system starting in a lattice configuration
remains solid and a system starting from a liquid configuration
remains liquid within the length of the runs. The Lindemann
criterion for classical melting locates the transition at the
temperature at which $\gamma\simeq 0.15$. From the result in the
table the fcc--liquid transition temperature should be located
between 1000\,K and 1500\,K. Previous investigation of such a
transition has been performed by Car--Parrinello Molecular
Dynamics \cite{kh96}. This study suggests that at $r_s=1$ the
structure of the system at $T=0K$ is hcp but for $T>$100\,K the
bcc structure is more favorable (as in this work, protons were
considered as classical particles). The melting temperature of the
bcc lattice has been estimated by the Lindemann criterion around
350\,K,  significantly lower than the present estimate. We are
presently investigating the system of 54 protons in order to study
the stability of the bcc structure and the melting transition
temperature.

\begin{table}
\caption{$r_s=1, N_p=N_e=32$, spin unpolarized}
\renewcommand{\arraystretch}{1.4}
\begin{center}
\begin{tabular}{|ccccccc|}
\hline\noalign{\smallskip}
$T(K)$&$K_{tot}$&$V_c$&$E_{tot}$  &$\sigma_E^2$&$P(Mbars)$&$\gamma$ \\
\noalign{\smallskip}
\hline
\noalign{\smallskip}
5000  &0.6241(2) &-0.7820(1) &-0.1579(2) &0.056(2) &21.72(2) &liquid \\
4000  &0.0620(2) &-0.7821(2) &-0.1619(1) &0.055(3) &21.35(1) &liquid \\
3000  &0.0616(1) &-0.7817(1) &-0.1662(2) &0.051(7) &20.93(1) &liquid \\
2000  &0.06122(6)&-0.7842(1) &-0.1702(1) &0.050(2) &20.588(6)&liquid \\
1500  &0.61113(7)&-0.7848(1) &-0.1737(1) &0.046(1) &20.374(6)&melted \\
1000  &0.60847(6)&-0.78372(8)&-0.17525(4)&0.0446(5)&20.181(3)&liquid \\
\noalign{\smallskip}
\hline
\noalign{\smallskip}
1000  &0.60894(5)&-0.78549(9)&-0.17655(7)&0.0416(5)&20.143(3)&0.137(4) \\
 700  &0.60787(3)&-0.78614(5)&-0.17817(8)&0.0402(6)&20.017(6)&0.109(2) \\
 500  &0.60811(3)&-0.78718(3)&-0.17913(5)&0.048(4) &19.985(3)&0.085(3) \\
 300  &0.60680(4)&-0.78686(2)&-0.18017(2)&0.042(3) &19.874(3)&0.083(1) \\
\noalign{\smallskip}
\hline
\noalign{\smallskip}
\end{tabular}
\end{center}
\label{tab:rs1n64}
\end{table}

\section{Conclusions and Outlook}

In this article we have discussed the CEIMC method, along with a
number of supporting developments to make it computationally
efficient. Using the penalty method, we have shown how it is
possible to formulate a classical Monte Carlo, with the energy
difference having statistical noise, without affecting the
asymptotic distribution of the protons. We have made significant
progress on several related issues: the computation of energy
differences, the development of wavefunctions that do not require
optimization of variational parameters and use of twist averaged
boundary conditions. We have applied the method to an important
many-body system, molecular and metallic hydrogen at high
pressure. We have shown that the method is feasible on current
multi-processor computers.

One of the advantages of QMC over DFT, in addition to higher
accuracy, is the different way basis sets enter. Single particle
methods usually work in a ``wave basis'', where the wave function
is expanded in plane waves or Gaussian orbitals.  In contrast QMC
uses a particle basis. A smooth basis (the trial wave function) is
indeed used within VMC, however, cusps and other features are
easily added without slowing down the computation. For this
reason, the bare Coulomb interaction can be easily treated, while
in LDA, typically a smooth pseudopotential is needed, even for
hydrogen, to avoid an excessive number of basis functions.

As shown in the example of twist-averaged boundary conditions, the
CEIMC method has further advantages when additional averages are to be
performed. In the present calculation, we assumed classical
protons for simplicity. Of course, quantum effects of the protons
are important and have been included in previous QMC and LDA
calculations.  But it is not hard to see that it is possible to do
path integrals of the nuclei within the penalty method for very
little increased cost over classical nuclei. A path integral
simulation creates a path of $M$ slices, with each slice at an
effective temperature of $MT$. We then need to perform $M$
separate electronic simulations, one for each slice. However, the
penalty method requires the error level to be approximately $k_B
T$. Then the required error level at each slice is $Mk_B T$, so
each of the $M$ separate QMC simulations need not be as accurate.
In contrast, for PI-LDA calculations, $M$ time slices will take
$M$ times as long.

Our impression is that the CEIMC method on this application  of
high pressure hydrogen has the same order of computational demands
as Car-Parrinello plane-wave methods: our results suggest that the
CEIMC method may turn out to be both more accurate and faster. The
processing power of current multi-processors is enough that
significant applications are being pursued, giving much more
accurate results for systems of hydrogen and helium including all
effects of electron correlation and quantum zero point motion. In
general, we expect the CEIMC method to be most useful when there
are additional averages to be performed perhaps due to disorder or
quantum effects. On the other hand DFT methods are more efficient
for optimizing molecular geometries where the existing functional
are known to be locally accurate or for dynamical studies outside
the scope of CEIMC.

Tests for non-hydrogenic systems are needed to find the
performance of the algorithms on a broader spectrum of
applications. The use of pseudopotentials within QMC to treat
atoms with inner core is well tested. What is not clear is how
much time will be needed to generate trial functions, and to
reduce the noise level to acceptable limits. Also interesting is
to develop a dynamical version of CEIMC, i.e. CEIMD.  Many of the
techniques discussed here  such as twist averaging, advanced trial
functions and energy difference methods are immediately
applicable. However, while it is possible within MC to allow
quantum noise, it is clear that the noise level on the forces must
be much smaller, since otherwise the generated trajectories will
be quite different. The effect of the quantum noise, in adding a
fictitious heat bath to the classical system, may negate important
aspects of the rigorous approach we have followed. One possible
approach is to locally fit the potential surface generated within
QMC to a smooth function, thereby reducing the noise level.
Clearly, further work is needed to allow this next step in the
development of microscopic simulation algorithms.

\section*{Acknowledgments}
This work has been supported by NSF DMR01-04399 and the
computational facilities at NCSA Urbana (Illinois-USA) and at CINES, Montpellier (France).
Thanks to M. Marechal for extending the facilities of CECAM.
C.P. acknowledges financial support from CNRS and from ESF through the SIMU program.
This work has been supported by the INFM Parallel Computing Initiative.

%


\begin{thebibliography}{8.}
\addcontentsline{toc}{section}{References}

\bibitem{metropolis53}
N.~Metropolis, A.~W. Rosenbluth, M. N. Rosenbluth, A. H. Teller, and E.~Teller: J. Chem. Phys. \textbf{21}, 1087 (1953)

\bibitem{parr89}
R.~G. Parr and W.~Yang.
\emph{Density Functional Theory of Atoms and Molecules}, Oxford, 1989.

\bibitem{car85}
R.~Car and M.~Parrinello: Phys. Rev. Lett. \textbf{55}, 2471 (1985)

\bibitem{payne92}
M.~C. Payne, M.~P. Teter, D.~C. Allan, T.~A. Arias, and J.~D. Joannopoulos: Rev. Mod. Phys \textbf{64}, 1045 (1992)

\bibitem{marx96}
D.~Marx and M.~Parrinello: J. Chem. Phys. {\bf 104}, 4077 (1996)

\bibitem{sprik00}
M.~Sprik: J. Phys.: Condens. Matter \textbf{12}, A161 (2000)

\bibitem{tuckerman00}
M.~E. Tuckerman and G.~J. Martyna: J. Phys. Chem. B \textbf{104}, 159 (2000)

\bibitem{foulkes}
W.~M.~C. Foulkes et al.: Rev. Mod. Phys. \textbf{73}, 33 (2001)

\bibitem{hammond94}
B.~L. Hammond, Jr. W.~A.~Lester, and P.~J. Reynolds.
\emph{Monte Carlo Methods in Ab Initio Quantum Chemistry},
World scientific lecture and course notes in chemistry, (World Scientific, Singapore, 1994)

\bibitem{anderson95}
J.~B. Anderson:`Exact quantum chemistry by {Monte Carlo} methods'
In  {\em Quantum Mechanical Electronic
  Structure Calculations with Chemical Accuracy}, ed. S.~R. Langhoff, (Kluwer Academic, 1995)

\bibitem{ceperley96}
D.~M. Ceperley and L.~Mitas:`Quantum Monte Carlo methods in chemistry'. In
\emph{ Advances in Chemical Physics} ed. by  I.~Prigogine and S.~A. Rice, (Wiley and Sons, 1996)

\bibitem{grossman95}
J.~C. Grossman, L.~Mitas, and K.~Raghavachari: Phys. Rev. Lett. \textbf{75}, 3870 (1995)

\bibitem{dewing00b}
M.~Dewing: Monte Carlo Methods: Application to hydrogen gas and hard spheres.
PhD thesis, University of Illinois at Urbana-Champaign (2000).
\newblock Available as arXiv:physics/0012030.

\bibitem{hubbard84}
W.~B. Hubbard and D.~J. Stevenson: `Interior structure'.
In {\em Saturn}, ed. by T.~Gehrels and M.~S. Matthews (University of Arizona Press, 1984)

\bibitem{stevenson88}
D.~J. Stevenson:`The role of high pressure experiment and theory in our understanding
of gaseous and icy planets'. In {\em Shock waves in condensed matter},
ed. by S.~C. Schmidt and N.~C. Holmes (Elsevier, 1988)

\bibitem{militzer00a}
B.~Militzer and E.~L. Pollock: Phys. Rev. E \textbf{61}, 3470 (2000)

\bibitem{ca87}
D.~M. Ceperley and B.~J. Alder: Phys. Rev. B \textbf{36}, 2092 (1987)

\bibitem{natoli93}
V.~Natoli, R.~M. Martin, and D.~M. Ceperley: Phys. Rev. Lett. \textbf{70}, 1952 (1993)

\bibitem{natoli95}
V.~Natoli, R.~M. Martin, and D.~M. Ceperley: Phys. Rev. Lett. \textbf{74}, 1601 (1995)

\bibitem{dewing}
M. Dewing and D.~M. Ceperley: `Methods for Coupled Electronic-Ionic Monte Carlo'. In:
\emph{Recent Advances in Quantum Monte Carlo Methods, II}, ed.
     by S. Rothstein (World Scientific, Singapore), submitted Jan 2001.

\bibitem{reynolds82}
P.~J. Reynolds, D.~M. Ceperley, B.~J. Alder, and W.~A. Lester: J. Chem. Phys. \textbf{77}, 5593 (1982)

\bibitem{ceperley99}
D.~M. Ceperley and M.~Dewing: J. Chem. Phys. \textbf{110}, 9812 (1999)

\bibitem{ceperley95}
D.~M. Ceperley: Rev. Mod. Phys. \textbf{67}, 279 (1995)

\bibitem{chep02}
D.~M. Ceperley, M. Holzmann, K. Esler and C. Pierleoni: Backflow Correlations for Liquid Metallic Hydrogen,
to appear.

\bibitem{kwon2d}
Y. Kwon, D.M. Ceperley and R. M. Martin: Phys. Rev. \textbf{B 48}, 12037 (1993)

\bibitem{kwon3d}
Y. Kwon, D.M. Ceperley and R. M. Martin: Phys. Rev. \textbf{B 58}, 6800 (1998)

\bibitem{cep84}
D. M. Ceperley and B. J. Alder: J. Chem. Phys. \textbf{81}, 5833 (1984)

\bibitem{ogitsu00}
T.~Ogitsu: `MP-DFT (multiple parallel density funtional theory) code, (2000)
http://www.ncsa.uiuc.edu/Apps/CMP/togitsu/MPdft.html.

\bibitem{silvera78}
I.~F. Silvera and V.~V. Goldman: J. Chem. Phys. \textbf{69}, 4209 (1978)

\bibitem{diep00a}
P.~Diep and J.~K. Johnson: J. Chem. Phys. \textbf{112}, 4465 (2000)

\bibitem{diep00b}
P.~Diep and J.~K. Johnson: J. Chem. Phys. \textbf{113}, 3480 (2000)

\bibitem{kolos64}
W.~Kolos and L.~Wolniewicz: J. Chem. Phys. \textbf{41}, 1964.

\bibitem{militzer00b}
B.~Militzer and D.~M. Ceperley: Phys. Rev. Lett. \textbf{85}, 1890 (2000)

\bibitem{militzer-thesis00}
B.~Militzer.
\newblock {\em Path Integral {Monte Carlo} Simulations of Hot Dense Hydrogen}.
\newblock PhD thesis, University of Illinois at Urbana-Champaign, 2000.

\bibitem{holmes95}
N.~C. Holmes, M.~Ross, and W.~J. Nellis: Phys. Rev. B \textbf{52}, 15835 (1995)

\bibitem{saumon91}
D.~Saumon and G.~Chabrier: Phys. Rev. A \textbf{44}, 5122 (1991)

\bibitem{saumon92}
D.~Saumon and G.~Chabrier: Phys. Rev. A \textbf{46}, 2084 (1992)

\bibitem{saumon95}
D.~Saumon, G.~Chabrier, and H.~M. {Van Horn}: Astrophys. J. Sup. \textbf{99}, 713 (1995)

\bibitem{nellis83}
W.~J. Nellis, A.~C. Mitchell, M.~van Theil, G.~J. Devine, R.~J Trainor, and
  N.~Brown: J. Chem. Phys. \textbf{79}, 1480 (1983)

\bibitem{hohl93}
D.~Hohl, V.~Natoli, D.~M. Ceperley, and R.~M. Martin: Phys. Rev. Lett. \textbf{71}, 541 (1993)

\bibitem{kohanoff97}
J.~Kohanoff, S.~Scandolo, G.~L. Chiarotti, and E.~Tosatti: Phys. Rev. Lett. \textbf{78}, 2783 (1997)

\bibitem{lzc01}
C. Lin, F. H. Zong and D. M. Ceperley: Phys. Rev. E \textbf{64}, 016702 (2001)

\bibitem{helium3-92}
D. M. Ceperley; Phys. Rev. Lett. \textbf{69}, 331 (1992)

\bibitem{pcbm94}
C. Pierleoni, B. Bernu, D. M. Ceperley and W. R. Magro: Phys. Rev. Lett. \textbf{73}, 2145 (1994);
W. R. Magro, D. M. Ceperley, C. Pierleoni, and B. Bernu: Phys. Rev. Lett. \textbf{76}, 1240 (1996)

\bibitem{como95}
D. M. Ceperley: `Path integral Monte Carlo methods for fermions'. In
\emph{Monte Carlo and Molecular Dynamics of Condensed Matter Systems},
ed. by K. Binder and G. Ciccotti (Editrice Compositori, Bologna, Italy, 1996)

\bibitem{kh96}
J. Kohanoff and J. P. Hansen: Phys. Rev. E \textbf{54}, 768 (1996)

\end{thebibliography}
\end{document}